\documentclass{evn2002}
\usepackage{graphicx}
\begin{document}
   \title{Compact Intraday Variable Radio Cores: New Observational Approaches}

   \author{L. Fuhrmann, G. Cim\`o,  T.P. Krichbaum, T. Beckert, 
           A. Kraus, A. Witzel, J. A. Zensus}

   \institute{Max-Planck-Institut f\"ur Radioastronomie, Bonn}

   \abstract{The evidence for refractive interstellar scintillation (RISS) 
being the main cause for rapid intraday variations (Intraday Variability, IDV) 
in Quasars and BL Lacs has recently become stronger.
If IDV is still a complex composition of extrinsic and source intrinsic effects,
the intrinsic part of the IDV pattern should show up in the millimeter and 
sub-millimeter regime due to the frequency dependence of RISS. Hence, observations 
at higher frequencies are essential in order to exclude RISS as the sole cause 
of IDV. Here we report on our new attempt to search for rapid variations at much 
higher frequencies. In addition, the possibility of a direct detection of the 
scattering screen in front of IDV sources will be discussed. Our recent line 
observations towards a few IDV sources lead to the first detection of a      
high latitude molecular cloud in front of an intraday variable radio core. 
}
    \authorrunning{Fuhrmann et al.}
    \maketitle
%

\section{Introduction}
There is still a hot debate about the origin of short term radio variations in total 
as well as polarized flux of flat spectrum Quasars and BL Lacs. Such variations on time 
scales of hours to days (IntraDay Variability, IDV) are common in this type of objects 
and reveal the tiny dimensions of the emitting region. The cause of the variations seen 
in these sources is currently controversial with claims being made for either: 1) a 
source-intrinsic or 2) extrinsic origin due to scattering in the interstellar medium (ISM)
(e.g. Wagner \& Witzel 1995 and ref. therein). Observational findings suggest, that IDV is 
a complicated mixture of both effects (Krichbaum et al. 2002).
Due to the involved small source sizes, refractive interstellar scintillation (RISS) must 
play an important role in the cm-radio regime, while intrinsic variations require extreme
high Doppler boosting or special source geometries.\\
In this paper we will present our new observational approaches: In the first part we will
report on the attempt to search for IDV at much higher frequencies than presently done 
(up to 345 GHz). In the second part we will introduce the possibility of a direct search 
for the required scattering screen and show first results of line observations towards IDVs.

\section{Search for Intrinsic IDV: High Frequency Observations}

During 48 hours of observations in September 1998 we found significant 
variations in the IDV-source 0716+714 at 32 GHz with 20\% peak-to-peak
variations (Krichbaum et al. 2002, Kraus et al. in prep.). 
This is the first detection of IDV at higher frequencies. Such mm-IDV is of 
particular interest due to the frequency dependence of refractive interstellar 
scintillation. Towards higher frequencies scintillation vanishes 
and the intrinsic part of IDV should show up in the millimeter and sub-millimeter 
regime. Hence, observations of IDV sources at such high frequencies provide the most 
direct observational test for RISS and will help to disentangle between source 
intrinsic and scintillation induced IDV.\\
Motivated by the results from September 1998, we performed new high frequency 
observations with Effelsberg and the HHT during the last months.

\subsection{Effelsberg Observations at 32 GHz}
In April 2002 we observed a small sample of IDV sources during a 40 hours 
session with the Effelsberg telescope using the new module of the 32 GHz correlation 
receiver. As in the past, we measured using the method of
cross-scans, with the number of subscans matched to the source brightness. A dense time 
coverage, with frequent observations of secondary calibrators in a duty cycle of 1-2 
measurements per source and per 1.5 hours has been done. The data reduction is still in progress. 
Up to now, the first inspection of the data does not show significant variations 
(compared to our secondary calibrators). Perhaps this is due to the not optimal weather conditions 
during the run (periods of clouds and even rain), but a further and more detailed data 
reduction has to be performed.    
 
\subsection{HHT Observations at 345 GHz}
In May 2002 we started a 3 week multifrequency campaign with the HHT
together with other radio as well as optical telescopes in order to search for 
short term variations at 345 GHz in flat spectrum radio sources (Cim\`o et al. in prep.). 
The data reduction is still in progress, but preliminary results may reveal 
IDV in at least a few sources. Here, the quasar 0716+714 is the most promising 
candidate and seems to be highly variable at such high frequencies during the campaign
(see also Cim\`o et al., this conference).

\section{Search for the Scattering Screen in Front of IDVs}
Another new observational approach is the attempt to directly detect the scattering screen 
in the foreground of intraday variables. The presence of a scattering screen in the Galaxy 
responsible for RISS has been suggested many years ago. Until now, only indirect hints do exist and no 
attempts has been performed in order to directly detect such medium in front of IDV-sources.
A direct detection via spectral line observations will help to identify RISS with the location of 
particular features in the local ISM (LISM). In addition, physical parameters deduced from 
such observations would give new input for the model of interstellar scintillation. \\
In order to perform such search we have to deal with the following questions: 1) which component 
of the ISM could serve as scattering material? 2) where is its location in the Galaxy? and 3) 
do we have observational evidence for the material? Scintillation of extragalactic radio sources 
on timescales of about 1 day requires free electrons (in a layer of a few parsec thickness)
at a relatively nearby position between us and the source. Such turbulent (clumpy) plasma produces 
a scintillation pattern onto the earth orbital plane by focusing and defocusing the radiation 
from the source. As the earth moves through this pattern, we observe temporal variations
of the flux emitted by the extragalactic object. In order to display such variations the source 
size $\theta$ must be of the order of the scattering size of the medium, which is defined by the Fresnel 
scale $\theta_{F}$. For source sizes larger than $\theta_{F}$, the scintillations 
are quenched (see Beckert et al., this proceedings). The observed variability timescale is related to 
the velocity of the earth $v$ 
relative to the scattering medium and the distance $d$ to the screen by $t\sim\theta d/v$. For typical 
velocities of 25 km/s and time scales of about one day the scattering plasma should not be further 
away than 100--150\,pc. Observational findings suggest a very inhomogeneous medium, which varies on 
angular scales not larger than a few degrees on the sky (Fuhrmann et al. 2002). Thus, we have to 
search for a somehow diffuse ionized component (with a certain amount of electron densities) of the 
ISM within a radius of $\sim 150$\,pc of the Sun. \\
In view of the required ionized material, several forms of such gas do exist. 
Since an extended, diffuse material as the structure of the general ISM or large extended
envelops of more compact structures are required, one may expect the warm ionized medium (WIM) or 
the extended low-density warm ionized medium (ELDWIM) to be connected with scintillation.
Such medium has been identified through detection of radio recombination lines (RRLs) at several 
positions free of discrete continuum sources already 30 years ago (e.g. Gottesmann \& Gordon 1970). 
This may give us the possibility to observe directly the postulated scattering screen in front 
of IDV sources via RRL observations.\\  
The ISM surrounding the solar system within a few hundred parsecs is known to contain several major 
features in the form of clouds, bubbles, supernova shells, star forming regions and molecular as well 
as HI clouds. While the Sun is thought to reside in the hot, low-density ($n_{e}\sim0.005\,cm^{-3}$) 
X-ray emitting Local Bubble (LB) with a radius of $\sim\,100$\,pc (e.g. Cox \& Reynolds 1987), the LB 
itself is embbeded into four large, almost circular rings, the Radio Loops I to IV (Haslam, Khan \& Meabur
n 1971). The sun lies more local inside the warm ($T\sim\,7000$\,K), diffuse and partially ionized
($n_{e}\geq\,0.12\,cm^{-3}$) local interstellar cloud (LIC) ($d\sim$ a few pc) (Lallement \& Bertin 1992).
The LIC extends for several parsecs in most directions and is part of the "local fluff" (Frisch 1995). 
This complex of warm, nearby clouds located either insight or at the edge of the LB is originated from 
the Sco-Cen star forming region in Loop I. In addition, the LB is surrounded by other well-known examples 
of nearby bubbles, star forming regions, molecular clouds and shells of SN remnants suitable to either 
contain or produce "scattering" material within a radius of 100--450\,pc.
Thus, it is quite plausible to think about such inhomogeneous clouds or shells at this distance with the 
required amount of ionized material and clumpyness as origin of the RISS screen.\\
Present galactic surveys of e.g. dust, HI, HII, H$\alpha$ provide additional information and enable 
us to search directly for observational evidence of material in each individual source direction. From 
this we can deduce that nearly all our IDV-sources lie in regions of enhanced emission, while often longterm-
or non-variable sources are located in regions of less emission or even in ``holes''. In addition, a 
high latitude molecular cloud (HLC) at a distance of only $\sim100$\,pc and separated only a few degrees from 
0917+624 and 0954+658 has been found (Magnani \& Blitz \& Mundy 1985). In this case, one could 
think of the ionized screen as extended envelop around the molecular cloud.\\
Using the H$\alpha$ data from the WHAM-survey (Haffner et al. in prep.) one can derive lower limits for 
recombination line temperatures $T_{L}$ in each individual source direction. Such temperatures are very low
and will be hard to detect. However, lines of other interstellar material can be used to search for the 
RISS-screen.

   \begin{figure*}
   \centering
   \hspace{1.5cm}
   \includegraphics[width=16cm,angle=0] {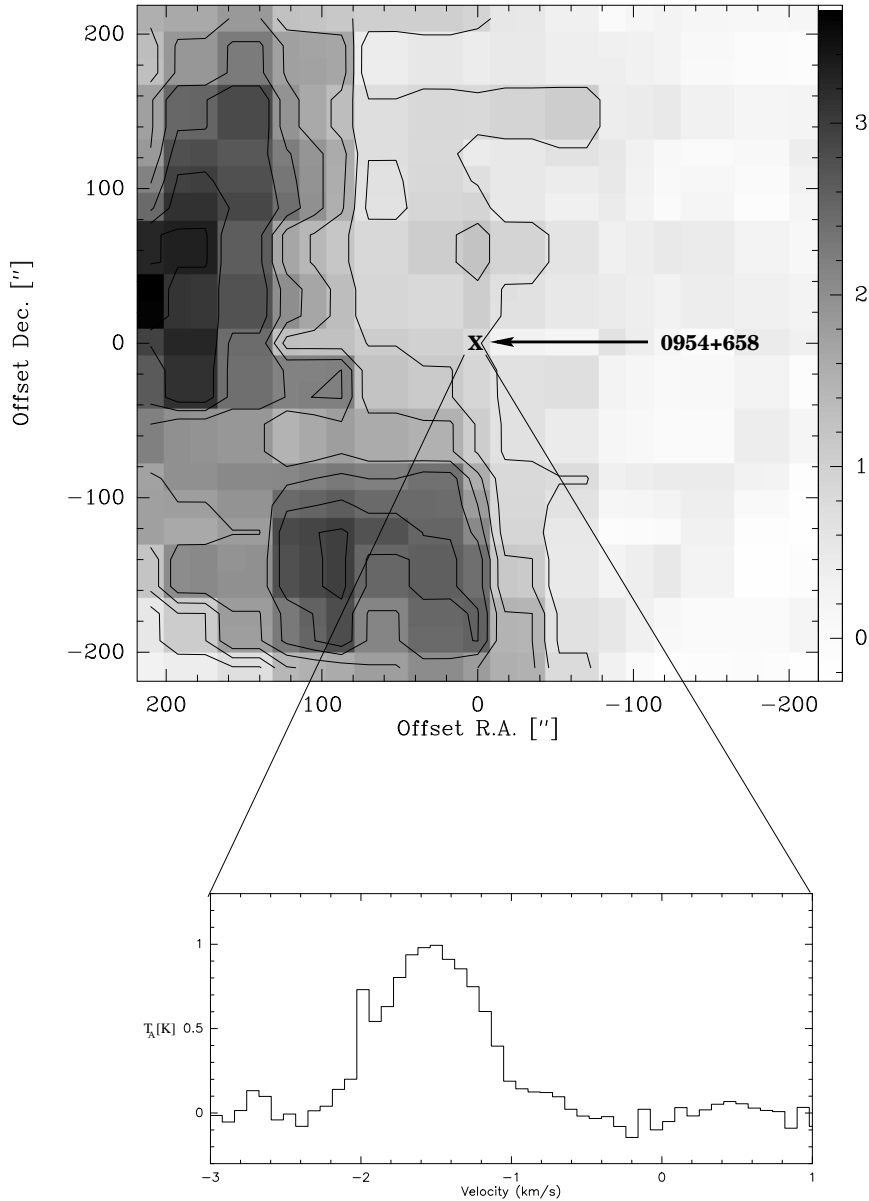}
   \caption{The detection of a CO cloud in front of the BL Lac object 0954+658 is displayed. Upper map:
            the cloud is located eastern of 0954+658 and extends in the north-south direction with 
            a minimum extension of 440$''$. The position of 0954+658 is marked with a cross. 
            Contour levels are: 20, 30, 40, 50, 60, 70, 80, 90 and $100\%$ of the maximum line emission  
            (lowest contour = $5\sigma$). The IDV source is located behind the outer CO-shell of the cloud.
            Lower panel: Spectrum at the position of 0954+658. The Intensity of the line is $T_{A}=1.005K$
            at a LSR velocity of $v=-1.54\,km/s$.}
            
    \end{figure*}
   
\section{Detection of CO in Front of 0954+658}
During our 345 GHz monitoring at the HHT in May 2002 we were able to perform additional observations: 
a few spectral lines towards IDVs has been observed in order to search for the nearby screen. 
Due to the existence of a CO-HLC near the IDV sources 0917+624 and 0954+658 we first used the 230 GHz SIS-receiver 
and were able to detect the CO(2-1)-line in the direction of the BL Lac object 0954+658. In Figure 1, a raster 
map of $7'\times 7'$ with a total observing time of $\sim15$\,h is displayed. The map is centered on the 
position of 0954+658. A CO cloud eastern of the IDV source is seen extending in the north-south direction.
0954+658 is located in the outer western part of the cloud.    
\section{Molecular Clouds as Harbour of the Scattering Material in Front of IDVs?} 
The larger HLC located near 0954+658 and 0917+624 is part of the Ursa Major clouds (Heithausen et al. 1993). 
Assuming the CO cloud in Figure 1 belongs also to this cloud complex, the distance to our detected object 
should not much further away than $\sim100$\,pc. This fits nicely with the expected distance coming 
from RISS for typical time scales seen in this source of $\sim 10$h. 
Since we have to search for ionized material, further spectral line observations have to be done (e.g. HCO+,
RRL). 0954+658 resides behind the outer shell of the cloud in Figure 1. It seems quite plausible to think about 
an ionized shell or envelop at this position surrounding the cloud and acting as ``shield'' against the interstellar 
radiation field. Additional observations of HCO+ and H$30\alpha$ on the line of sight to the BL Lac~object 
have been done and are still inconclusive, but with hints of weak lines (Fuhrmann et al. in prep.). 
Since such lines are expected to be extremely weak, new observations with higher sensitivity are  
necessary (IRAM 30m). \\
0954+658 is of particular interest. This source was the first object in which a variability pattern 
resembling an extreme scattering event (ESE) was seen (Fiedler et al. 1987). An ESE occurs, when an 
isolated inhomogeneity (plasma lens) in the ISM passes through the line of sight to the extragalactic 
object. Such event causes ray path distortions and leads to dramatic flux density variations. In March 2000
an even more rapid ESE in 0954+658 has been observed (Cim\`o et al. 2002). In view of our recent findings 
such event suggests a very clumpy and extremely turbulent outer envelop or boundary layer around the CO cloud
in Figure 1.\\
In order to identify CO clouds as origin of the scattering material responsible for RISS, present CO surveys 
provide additional information (Hartmann, Magnani \& Thaddeus, 1998). Comparing the positions of all our 
IDVs in the northern galactic hemisphere with the positions of HLCs it turns out that extremely often IDVs 
are located near ($\leq 5\degr-10\degr$) CO clouds or cloud complexes. This has to be investigated in more 
detail and will be discussed elsewhere (Fuhrmann et al. in prep.).

\section{Summary}
During the first half of 2002 we started two new observational approaches: a search for high frequency IDV 
(32 GHz and 345 GHz) and the attempt to detect directly the possible foreground screen via spectral line 
observations. Since in both cases the data reduction is still in progress, preliminary results show: 
1) if confirmed, at least the BL Lac object 0716+714 seems to show sub-mm IDV. This can not be explained 
by RISS and must be due to intrinsic mechanism. 2) a CO cloud in the direction of 0954+658. Such high 
latitude clouds could serve as the origin of the scattering material producing interstellar scintillation. 
This has to be confirmed by additional investigations of ionized material, like ionized HCO or the detection of 
RRLs. A further option could be faraday rotation measures on the line of sight to the source compared with those 
measured a few degrees away.

\end{document}